\documentclass[aps,prl,superscriptaddress,amsmath,amssymb,twocolumn,
nofootinbib,
]{revtex4-2}
\usepackage[utf8]{inputenc}
\usepackage{graphicx}% Include figure files
\usepackage{dcolumn}% Align table columns on decimal point
\usepackage{bm}% bold math
\usepackage{amsmath}
\usepackage[normalem]{ulem}

\usepackage{color}

\newcommand{\ele}[2]{$^{#1}$#2}
\newcommand{\oiso}{$^{16}$O$/^{17}$O~}
\newcommand{\ciso}{$^{12}$C$/^{13}$C~}
\newcommand{\op}{$^{16}$O$(p,\gamma)^{17}$F~}
\newcommand{\oopg}{$^{17}$O$(p,\gamma)^{18}$F~}
\newcommand{\oopa}{$^{17}$O$(p,\alpha)^{14}$N~}

\newcommand{\pg}{\ensuremath{(\mathrm{p},\gamma)}}

\newcommand{\pa}{\ensuremath{(\mathrm{p},\alpha)}}

\newcommand{\ion}[3]{\ensuremath{{}^{#1}\mathrm{#2}^{#3}}}
\newcommand{\nuc}[2]{\ion{#1}{#2}{}}
\newcommand{\Bp}{\ensuremath{\beta^{+}}}
\newcommand{\react}[5]{\ensuremath{\nuc{#1}{#2}#3\nuc{#4}{#5}}}
\newcommand*{\fullOst}{\react{16}{O}{\pg}{17}{F}}

\newcommand{\Ep}{\ensuremath{E_\mathrm{p}}}

\newcommand{\dctof}{DC$\rightarrow$495}
\newcommand{\dctog}{DC$\rightarrow$0}

\newcommand{\CeBr}{$\mathrm{CeBr}_\mathrm{3}$}

\newcommand*{\degs}{$^{\circ}$}

\begin{document}

\preprint{APS/123-QED}
%TC:ignore
\title{\textbf{The $^{16}$O(p,$\gamma$)$^{17}$F reaction and the C and O isotopic ratios in red giant branch stars} 
}% 

%\author{Ann Author}
% \altaffiliation[Also at ]{Physics Department, XYZ University.}%Lines break automatically or can be forced with \\
%\author{Second Author}%
% \email{Contact author: Second.Author@institution.edu}
%\affiliation{%
% Authors' affiliations\\
%  Include all institutions where the work was conducted: department or division, institution, city, state (if relevant), and country, in this order.
%}%
\author{D.~Robb}
 \affiliation{SUPA, School of Physics and Astronomy, University of Edinburgh, EH9 3FD Edinburgh, United Kingdom}

\author{P.~McGlynn}
 \affiliation{Facility for Rare Isotope Beams, Michigan State University, East Lansing, MI 48824, USA}
 
\author{M.~Aliotta}
 \affiliation{SUPA, School of Physics and Astronomy, University of Edinburgh, EH9 3FD Edinburgh, United Kingdom}

\author{C.~Ananna}
\thanks{Presently at Universit\`a del Salento}
 \affiliation{Universit\`a degli Studi di Napoli ``Federico II'', 80125 Naples, Italy}
 \affiliation{Istituto Nazionale di Fisica Nucleare, Sezione di Napoli, 80125 Naples, Italy} 
 
\author{F.~Barile}
 \affiliation{Università degli Studi di Bari ``A. Moro'', 70125 Bari, Italy}
 \affiliation{Istituto Nazionale di Fisica Nucleare, Sezione di Bari, 70125 Bari, Italy}

\author{L.~Barbieri}
 \affiliation{SUPA, School of Physics and Astronomy, University of Edinburgh, EH9 3FD Edinburgh, United Kingdom}
 
\author{D.~Bemmerer}
 \affiliation{Helmholtz-Zentrum Dresden-Rossendorf, 01328 Dresden, Germany}

\author{A.~Best}
 \affiliation{Universit\`a degli Studi di Napoli ``Federico II'', 80125 Naples, Italy}
 \affiliation{Istituto Nazionale di Fisica Nucleare, Sezione di Napoli, 80125 Naples, Italy} 

\author{R.~Biasissi}
 \affiliation{Dipartimento di Fisica, Università degli Studi di Padova, 35131 Padova, Italy}
 \affiliation{Istituto Nazionale di Fisica Nucleare, Sezione di Padova, 35131 Padova, Italy} 
 
\author{A.~Boeltzig}
 \affiliation{Helmholtz-Zentrum Dresden-Rossendorf, 01328 Dresden, Germany}

\author{R.~Bonnell}
 \affiliation{SUPA, School of Physics and Astronomy, University of Edinburgh, EH9 3FD Edinburgh, United Kingdom}
 
\author{C.~Broggini}
 \affiliation{Istituto Nazionale di Fisica Nucleare, Sezione di Padova, 35131 Padova, Italy}
 
\author{C.G.~Bruno}
 \affiliation{SUPA, School of Physics and Astronomy, University of Edinburgh, EH9 3FD Edinburgh, United Kingdom}

\author{A.~Caciolli}
 \affiliation{Dipartimento di Fisica, Università degli Studi di Padova, 35131 Padova, Italy}
 \affiliation{Istituto Nazionale di Fisica Nucleare, Sezione di Padova, 35131 Padova, Italy} 
 
\author{M.~Campostrini}
 \affiliation{Laboratori Nazionali di Legnaro, 35020 Legnaro, Italy}
 
\author{F.~Casaburo}
 \affiliation{Università degli Studi di Genova, 16146 Genova, Italy}
 \affiliation{Istituto Nazionale di Fisica Nucleare, Sezione di Genova, 16146 Genova, Italy}
 
\author{F.~Cavanna}
 \affiliation{Istituto Nazionale di Fisica Nucleare, Sezione di Torino, 10125 Torino, Italy}

\author{G.F.~Ciani}
 \affiliation{Istituto Nazionale di Fisica Nucleare, Sezione di Bari, 70125 Bari, Italy}
 
\author{T.~Chillery}
 \affiliation{Istituto Nazionale di Fisica Nucleare, Laboratori Nazionali del Gran Sasso, 67100 Assergi, Italy}

\author{P.~Colombetti}
 \affiliation{Dipartimento di Fisica, Universit\`a degli Studi di Torino, 10125 Torino, Italy}
 \affiliation{Istituto Nazionale di Fisica Nucleare, Sezione di Torino, 10125 Torino, Italy}
 
\author{A.~Compagnucci}
 \affiliation{SUPA, School of Physics and Astronomy, University of Edinburgh, EH9 3FD Edinburgh, United Kingdom}

\author{P.~Corvisiero}
 \affiliation{Università degli Studi di Genova, 16146 Genova, Italy}
 \affiliation{Istituto Nazionale di Fisica Nucleare, Sezione di Genova, 16146 Genova, Italy}
 
\author{L.~Csedreki}
 \affiliation{HUN-REN Institute for Nuclear Research (ATOMKI), PO Box 51, H-4001 Debrecen, Hungary}

\author{T.~Davinson}
 \affiliation{SUPA, School of Physics and Astronomy, University of Edinburgh, EH9 3FD Edinburgh, United Kingdom}

 \author{R.~Depalo}
  \email{rosanna.depalo@unimi.it}
 \affiliation{Università degli Studi di Milano, 20133 Milano, Italy}
 \affiliation{Istituto Nazionale di Fisica Nucleare, Sezione di Milano, 20133 Milano, Italy}

\author{A.~Di~Leva}
 \affiliation{Universit\`a degli Studi di Napoli ``Federico II'', 80125 Naples, Italy}
 \affiliation{Istituto Nazionale di Fisica Nucleare, Sezione di Napoli, 80125 Naples, Italy}

\author{Z.~Elekes}
 \affiliation{HUN-REN Institute for Nuclear Research (ATOMKI), PO Box 51, H-4001 Debrecen, Hungary}

\author{F.~Ferraro}
 \affiliation{Istituto Nazionale di Fisica Nucleare, Laboratori Nazionali del Gran Sasso, 67100 Assergi, Italy}

\author{A.~Formicola}
 \affiliation{Istituto Nazionale di Fisica Nucleare, Sezione di Roma, Piazzale A. Moro 2, 00185 Roma, Italy}

\author{Zs.~F\"ul\"op}
 \affiliation{HUN-REN Institute for Nuclear Research (ATOMKI), PO Box 51, H-4001 Debrecen, Hungary}

\author{G.~Gervino}
 \affiliation{Dipartimento di Fisica, Universit\`a degli Studi di Torino, 10125 Torino, Italy}
 \affiliation{Istituto Nazionale di Fisica Nucleare, Sezione di Torino, 10125 Torino, Italy}

\author{R.\,M.~Gesu\`e}
 \affiliation{Gran Sasso Science Institute, 67100 L'Aquila, Italy}
 \affiliation{Istituto Nazionale di Fisica Nucleare, Laboratori Nazionali del Gran Sasso, 67100 Assergi, Italy}

\author{G.~Gosta}
 \affiliation{Università degli Studi di Milano, 20133 Milano, Italy}
 \affiliation{Istituto Nazionale di Fisica Nucleare, Sezione di Milano, 20133 Milano, Italy}
 
\author{Gy.~Gy\"urky}
 \affiliation{HUN-REN Institute for Nuclear Research (ATOMKI), PO Box 51, H-4001 Debrecen, Hungary}

\author{A.~Guglielmetti}
 \affiliation{Università degli Studi di Milano, 20133 Milano, Italy}
 \affiliation{Istituto Nazionale di Fisica Nucleare, Sezione di Milano, 20133 Milano, Italy}

\author{C.~Gustavino}
 \affiliation{Istituto Nazionale di Fisica Nucleare, Sezione di Roma, 00185 Roma, Italy}

\author{C.~Hebborn}
 \affiliation{CNRS - IJCLab, 91405 Orsay, France}

\author{G.~Imbriani}
 \affiliation{Universit\`a degli Studi di Napoli ``Federico II'', 80125 Naples, Italy}
 \affiliation{Istituto Nazionale di Fisica Nucleare, Sezione di Napoli, 80125 Naples, Italy}
 
\author{M.~Junker}
 \affiliation{Istituto Nazionale di Fisica Nucleare, Laboratori Nazionali del Gran Sasso, 67100 Assergi, Italy}

\author{M.~Lugaro}
 \affiliation{Konkoly Observatory, Research Centre for Astronomy and Earth Sciences (HUN-REN CSFK), 1121 Budapest, Hungary}
 
\author{J.~Marsh}
 \affiliation{SUPA, School of Physics and Astronomy, University of Edinburgh, EH9 3FD Edinburgh, United Kingdom}

\author{E.~Masha}
 \affiliation{Helmholtz-Zentrum Dresden-Rossendorf, 01328 Dresden, Germany}

\author{R.~Menegazzo}
 \affiliation{Istituto Nazionale di Fisica Nucleare, Sezione di Padova, 35131 Padova, Italy}

\author{D.~Mercogliano}
 \affiliation{Universit\`a degli Studi di Napoli ``Federico II'', 80125 Naples, Italy}
 \affiliation{Istituto Nazionale di Fisica Nucleare, Sezione di Napoli, 80125 Naples, Italy} 

\author{V.~Paticchio}
 \affiliation{Istituto Nazionale di Fisica Nucleare, Sezione di Bari, 70125 Bari, Italy}

\author{D.~Piatti}
 \affiliation{Dipartimento di Fisica, Università degli Studi di Padova, 35131 Padova, Italy}
 \affiliation{Istituto Nazionale di Fisica Nucleare, Sezione di Padova, 35131 Padova, Italy} 

\author{P.~Prati}
 \affiliation{Università degli Studi di Genova, 16146 Genova, Italy}
 \affiliation{Istituto Nazionale di Fisica Nucleare, Sezione di Genova, 16146 Genova, Italy}

\author{D.~Rapagnani}
 \affiliation{Universit\`a degli Studi di Napoli ``Federico II'', 80125 Naples, Italy}
 \affiliation{Istituto Nazionale di Fisica Nucleare, Sezione di Napoli, 80125 Naples, Italy} 
 
\author{V.~Rigato}
 \affiliation{Laboratori Nazionali di Legnaro, 35020 Legnaro, Italy}

\author{M.~Rossi}
 \affiliation{Istituto Nazionale di Fisica Nucleare, Sezione di Genova, 16146 Genova, Italy}

 \author{R.~Sariyal}
 \affiliation{Istituto Nazionale di Fisica Nucleare, Sezione di Torino, 10125 Torino, Italy}

\author{R.S.~Sidhu}
\thanks{Presently at University of Surrey}
 \affiliation{SUPA, School of Physics and Astronomy, University of Edinburgh, EH9 3FD Edinburgh, United Kingdom}

\author{J.~Skowronski}
 \affiliation{Dipartimento di Fisica, Università degli Studi di Padova, 35131 Padova, Italy}
 \affiliation{Istituto Nazionale di Fisica Nucleare, Sezione di Padova, 35131 Padova, Italy} 
 
\author{O.~Straniero}
 \affiliation{Istituto Nazionale di Astrofisica, Osservatorio Astronomico d'Abruzzo, Via Mentore Maggini, 64100, Teramo, Italy}
 \affiliation{Istituto Nazionale di Fisica Nucleare, Sezione di Roma, Piazzale A. Moro 2, 00185 Roma, Italy}

\author{T.~Sz\"ucs}
 \affiliation{HUN-REN Institute for Nuclear Research (ATOMKI), PO Box 51, H-4001 Debrecen, Hungary}

\author{S.~Turkat}
\thanks{Presently at TU Dresden}
 \affiliation{Dipartimento di Fisica, Università degli Studi di Padova, 35131 Padova, Italy}
 \affiliation{Istituto Nazionale di Fisica Nucleare, Sezione di Padova, 35131 Padova, Italy} 
 
\author{M.~Vagnoni}
 \affiliation{Universit\`a degli Studi della Campania “Luigi Vanvitelli,” 81100 Caserta, Italy}
 \affiliation{Istituto Nazionale di Fisica Nucleare, Sezione di Roma, Piazzale A. Moro 2, 00185 Roma, Italy}
 
\author{S.~Zavatarelli}
 \affiliation{Istituto Nazionale di Fisica Nucleare, Sezione di Genova, 16146 Genova, Italy}

\collaboration{LUNA Collaboration}\noaffiliation

%\date{}% It is always \today, today,
             %  but any date may be explicitly specified

\begin{abstract}
The \nuc{16}{O}/\nuc{17}{O} isotopic ratio is a widely used tracer of nucleosynthesis and mixing processes in red giant stars. Its value depends on the rates of the proton-capture reactions responsible for the production and destruction of \nuc{16}{O} and \nuc{17}{O} among which the \react{16}{O}{\pg}{17}{F} reaction plays a central role, as it involves both isotopes.
This work reports on a new measurement of the \react{16}{O}{\pg}{17}{F} cross section carried out at the Laboratory for Underground Nuclear Astrophysics (LUNA) over a center-of-mass energy range of 200–380~keV, a region previously unexplored by experimental studies. Two complementary techniques were employed: prompt gamma-ray detection and activation. The results obtained with the two techniques are consistent within experimental uncertainties. From the measured cross sections and existing literature data, we evaluate a new thermonuclear reaction rate, which is then used to demonstrate how simultaneous measurements of stellar carbon and oxygen isotopic ratios constrain the initial mass and the extent of the mixing events in red giant branch stars. In particular, we demonstrate how the $^{12}$C/$^{13}$C abundance ratio is primarily sensitive to the temperature T$_{\textrm{max}}$ attained during mixing, whereas the $^{16}$O/$^{17}$O abundance ratio is highly sensitive to the initial stellar mass M$_i$. Therefore, simultaneous measurements of these isotopic ratios can disentangle and place stringent constraints on M$_i$ and T$_{\textrm{max}}$, and the remaining uncertainty is now dominated by stellar abundance measurements rather than nuclear-physics inputs.
%The fit incorporates data from this work as well as previous measurements available in the literature. It is based on a two-body model in which the interaction between the proton and the \nuc{16}{O} nucleus is described using Woods–Saxon potentials. The resulting reaction rate agrees well with the most recent evaluation in the literature, although the associated uncertainty has increased to 8\%. 
\end{abstract}
%TC:endignore
%\keywords{Suggested keywords}%Use showkeys class option if keyword
                              %display desired
\maketitle

%\tableofcontents

%--------------------------------------------------------------
\section{\label{sec:intro}Introduction}
The isotopic ratios of carbon, nitrogen and oxygen measured in stellar atmospheres are sensitive tracers of the interplay between nuclear burning and mixing processes due to convection and/or other dynamical or secular instabilities. In particular, the photospheric \oiso ratio changes when the star reaches the red giant branch (RGB) and convective mixing extends from the stellar surface down to the layers where proton-capture reactions on oxygen isotopes are or have been previously active. Under such conditions, \ele{16}{O} is depleted, while the \ele{17}{O} is enhanced, leading to a net decrease in the isotopic ratio. Similarly, a reduced \ciso ratio indicates mixing into regions where proton-capture reactions on carbon isotopes are or have been previously active.  

Carbon isotopic ratios measured in low-mass RGB stars ($M\leq2$ M$_\odot$) are often lower than predicted by stellar models \cite{Abia2017}, supporting the existence of extra mixing in addition to the first dredge-up \cite{Lagarde2019,McCormick2023}.
%A wealth of observational evidence supports the existence of extra mixing in low-mass RGB stars ($M\leq2$ M$_\odot$) \cite{Lagarde2019,McCormick2023}, where observed carbon isotopic ratios are often lower than predicted by stellar models \cite{Abia2017}.
%. In these stars, the border of the convective envelope remains at temperatures well below the threshold needed to activate proton-capture reactions on CNO isotopes. Yet, observed carbon isotopic ratios are often lower than predicted by stellar models \cite{Abia2017}. 
Simultaneous measurements of both carbon and oxygen isotopic ratios offer a powerful tool for constraining the extent and efficiency of mixing events, provided that the nuclear reaction rates governing the production and destruction of the relevant isotopes are known with sufficient precision. 
Moreover those rates are needed to interpret oxygen isotopic ratios measured in presolar grains from AGB stars, which carry the fingerprint of mixing processes in these stars \cite{Lugaro2007, Lugaro2017}.

One key reaction in this context is \fullOst{}, which governs the destruction of \ele{16}{O} and produces \ele{17}{O} via \ele{17}{F} decay. At astrophysical temperatures (10 - 500~MK), this reaction proceeds via direct capture either to the $5/2^+$ ground state or to the $1/2^+$ excited state at 495~keV of \ele{17}{F}. A comprehensive evaluation of the available experimental data was conducted by Iliadis et al. in 2008 \cite{Iliadis2008}, who identified and, in some cases, corrected systematic issues in earlier measurements. The currently adopted reaction rate \cite{Iliadis2022, SF3} is based on a fit of existing data with a single-particle potential model combined with Bayesian statistical methods and carries an estimated uncertainty of 4.2\%. However, alternative fits to the data provide results differing by up to 10\% \cite{Tursunov2024}. Moreover, the only data set extending to centre-of-mass energies below 350~keV \cite{Hester1958} is affected by uncertainties as large as 50\%. In this work, we present a new measurement of the \fullOst{} cross section at centre-of-mass energies between 200 and 380~keV, thereby addressing a critical gap in the experimental data relevant to stellar nucleosynthesis and deep mixing in red giants.

%--------------------------------------------------------------
\section{\label{sec:experiment}Experiment}

%The experiment was carried out by bombarding solid \TaO{} targets \cite{Caciolli2012} with an intense ($\sim200$ $\mu$A) proton beam from the LUNA 400 kV linear accelerator \cite{Formicola2003} at Laboratori Nazionali del Gran Sasso. The targets were enriched to between 4 and 8\% in \nuc{18}{O} to allow characterisation by Nuclear Resonant Reaction Analysis (NRRA) \cite{LUNASolidTargets}, using the well-known resonance in \react{18}{O}{\pg}{19}{F} at \Ep{} = 151~keV \cite{Caciolli2012}. The measured energy loss at this energy was between 13 and 21~keV, and resonance scan were repeated periodically to monitor target degradation.

The experiment was carried out by bombarding solid oxygen targets with an intense ($\sim200$ $\mu$A) proton beam from the LUNA 400 kV accelerator \cite{Formicola2003} at Laboratori Nazionali del Gran Sasso (LNGS), with beam energies ranging from 230~keV to 400~keV. The targets were produced either through anodic oxidation of tantalum backings \cite{Caciolli2012}, or via reactive sputtering. Oxidised targets were enriched to between 4 and 8\% in \nuc{18}{O} to allow characterisation by Nuclear Resonant Reaction Analysis (NRRA) \cite{LUNASolidTargets}, using the well-known resonance in \react{18}{O}{\pg}{19}{F} at \Ep{} = 151~keV \cite{Caciolli2012}. The measured energy loss at this energy was between 13 and 21~keV, depending on the target, corresponding to $1.24\times10^{17}$ to $2.04\times10^{17}$ atoms/cm$^2$, respectively. The use of targets with different thicknesses provided a cross-check for possible systematic effects associated with the energy loss within the target. The NRRA was repeated periodically to monitor target degradation.

Two complementary experimental techniques were employed, both based on the detection of gamma rays. The first used a segmented BGO detector to perform an activation measurement exploiting the $\beta^+$ decay of \nuc{17}{F} (half-life of 64.4 s \cite{Brodeur2016}). Each data-acquisition run was carried out alternating periods of target irradiation and decay counting, the latter conducted with no beam on target.
The yield was determined by measuring the rate of the resulting annihilation radiation over time. The environmental background was effectively suppressed by selecting events corresponding to two 511~keV gamma rays detected in coincidence in diametrically opposed detector segments (180\degs{} apart),  \cite{Skowronski2023, c12c13_prl2023}. This approach enabled a total cross section measurement, free from uncertainties related to angular distribution or branching ratio effects.

In the second campaign, prompt gamma rays were detected directly using a HPGe detector positioned at 55\degs{} with respect to beam direction, together with two \CeBr{} scintillators placed nominally at 0\degs{} and 90\degs{}. This setup allowed for the determination of the direct-capture cross sections to the ground state and to the first excited state of \nuc{16}{O}, hereafter denoted \dctog{} and \dctof{}, respectively.

%Due to the low reaction Q-value (Q = 600.27~keV \cite{Wang2012}), the gamma rays emitted by the \react{16}{O}{\pg}{17}{F} reaction fall in a region where the background is dominated by environmental radioactivity. For this reason, 
Both setups employed passive shieldings consisting of lead and polyethylene to reduce environmental background \cite{Skowronski2023}. Combined with the deep underground environment of LNGS, this configuration suppressed both cosmic-ray-induced and environmental backgrounds \cite{Boeltzig2018, Skowronski2023}.%, enabling measurements at energies that are typically inaccessible in aboveground laboratories \cite{Aliotta2022}.

The absolute gamma-ray detection efficiencies for both experimental setups were measured using calibrated radioactive sources ($^{60}$Co, $^{133}$Ba and $^{137}$Cs) and the well-known resonance at \Ep = 278 keV in the $^{14}$N(p,$\gamma$)$^{15}$O reaction. These measurements were used to validate Monte Carlo simulations of the experimental configurations, implemented within the LUNA Geant4 \cite{Agostinelli2003} framework SimLUNA. For the activation campaign, the detection efficiency for 511~keV gamma-rays in coincidence was $29.6\%$. For the prompt-gamma experiment, the simulations were employed to determine the efficiency at the gamma-ray energies relevant to the $^{16}$O(p,$\gamma$)$^{17}$F reaction, yielding values between $0.78\%$ and $2.66\%$, depending on the detector and gamma-ray energy. The simulations were also used to quantify true-coincidence summing corrections. 

%Monte Carlo simulations of both experimental configurations were carried out using the LUNA Geant4 \cite{Agostinelli2003} framework SimLUNA. These simulations, along with measurements of calibrated radioactive sources ($^{60}$Co, $^{133}$Ba and $^{137}$Cs) \textcolor{red}{and the well-known resonance at \Ep = 278 keV in the $^{14}$N(p,$\gamma$)$^{15}$O reaction}, were used to determine the absolute detection efficiencies of the setups. \textcolor{red}{For the activation campaign, the detection efficiency for 511~keV gamma-rays in coincidence was $29.6\%$, while for the prompt-gamma campaign detection efficiencies ranged between $0.78\%$ and $2.66\%$, depending on the detector and on the gamma-ray energy.} %The simulation of the prompt gamma-ray setup was also used to investigate angular distribution effects and true-coincidence summing corrections.
%activation efficiency $$(29.6 \pm 1.5)\%$
% prompt efficiency  between $(0.78 \pm 0.04)$ and $(2.66 \pm 0.13)\%$
%--------------------------------------------------------------
\section{\label{sec:analysis}Data analysis}

%To calculate the \fullOst{} reaction yield from the activation data, the rate of back-to-back 511~keV coincidence events over time was fitted with the differential equation:

In the activation experiment, the \fullOst{} reaction yield was found as described in \cite{Skowronski2023}, by solving the differential equation:

\begin{equation}
    \label{eq:ProductionAndDecay}
    %\frac{dN}{dt}=KI(t) + \sum\limits_{i}\left(Y_{\rm{i}}I(t) - \lambda_{\rm{i}}N_{\rm{i}}(t)\right),
    \frac{dN_i}{dt}=Y_{\rm{i}}I(t) - \lambda_{\rm{i}}N_{\rm{i}}(t),
\end{equation}

\noindent where $N_{\rm{i}}(t)$ is the number of nuclei of species $i$ at time $t$, $\lambda_i$ is the decay constant of species $i$, $Y_{\rm{i}}$ is the yield of isotope $i$, expressed in counts per incident proton and $I(t)$ is the number of incident protons over time. The annihilation rate was then fitted with the following equation:

\begin{equation}
    \label{eq:AnnihilationRate}
    %\frac{dN}{dt}=KI(t) + \sum\limits_{i}\left(Y_{\rm{i}}I(t) - \lambda_{\rm{i}}N_{\rm{i}}(t)\right),
    A(t) = KI(t) + \sum\limits_{i} \lambda_{\rm{i}}N_{\rm{i}}(t),
\end{equation}

\begin{figure}[!htb]
    \centering
    \includegraphics[width=1.0\linewidth]{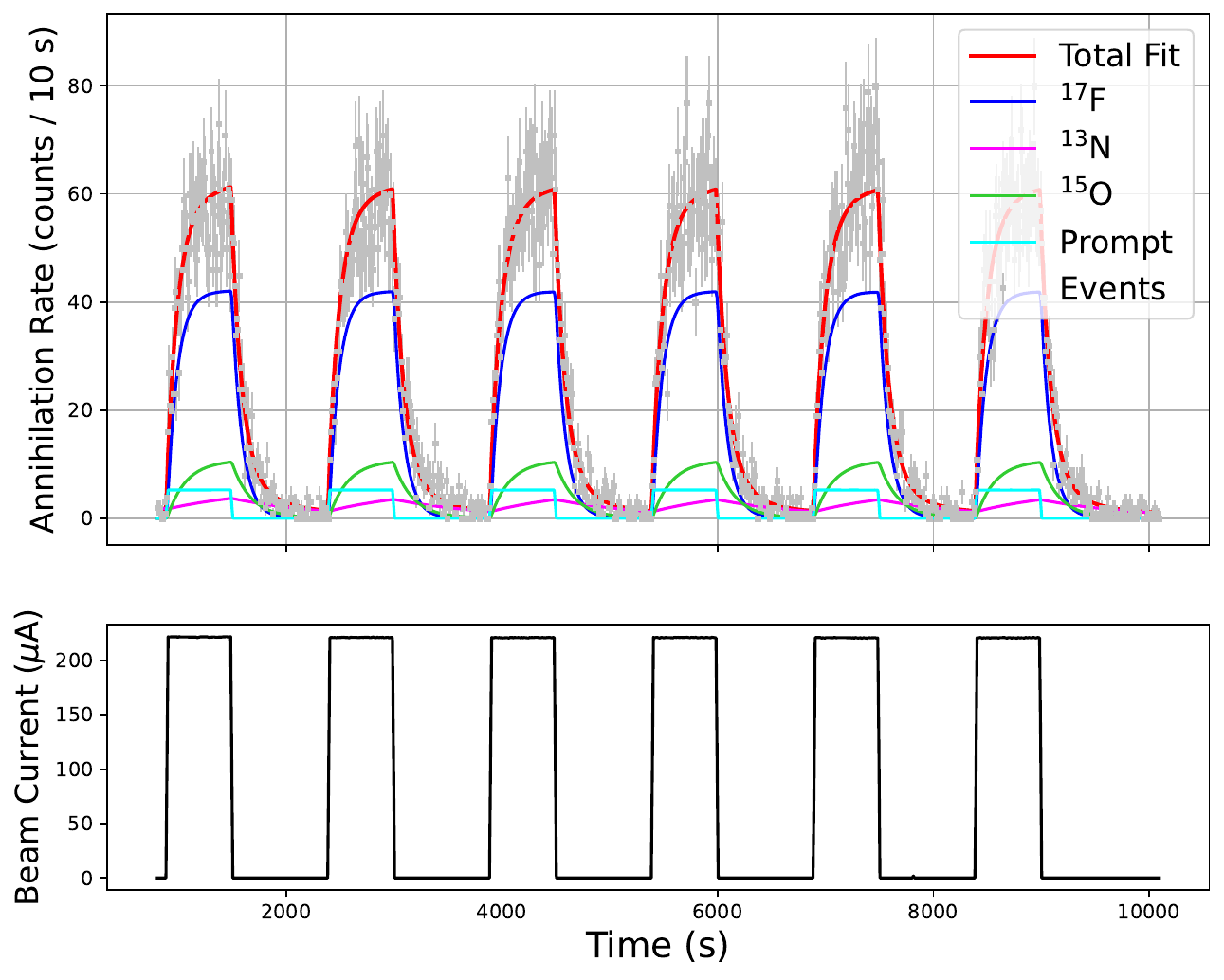}
    \caption{Top panel: sample annihilation-rate plot, at \Ep=330~keV, fitted with equation \ref{eq:AnnihilationRate}. The red line shows the total fit, while the contributions of the different terms in equation \ref{eq:ProductionAndDecay} are plotted with different colors. Bottom panel: Beam current as a function of time.}
    \label{fig:ActivationSpectrum}
\end{figure}

\noindent The term $KI(t)$ represents the contribution from prompt gamma rays, originating primarily from annihilation radiation caused by e$^-$e$^+$ pair production of high-energy gamma rays \cite{Skowronski2023}. This term was treated as a free parameter in the fitting procedure. %Its contribution becomes significant at beam energies corresponding to resonances in high Q-value reactions on contaminants, such as the \Ecm{} = 323~keV resonance in \react{19}{F}{\pag}{16}{O} \cite{Williams2021}.  
The sum term in equation \ref{eq:AnnihilationRate} accounts for the presence of \Bp{} emitters produced by proton-induced reactions on contaminants in the target, in addition to the expected \nuc{17}{F}. Two \Bp{} emitting isotopes were identified: \nuc{13}{N}, half-life 597.2 s \cite{Long2022}, produced by proton capture on \nuc{12}{C}; and \nuc{15}{O}, half-life of 122.3 s \cite{Burdette2020}, from proton capture on \nuc{14}{N}. The latter produced a detectable signal only at energies above the intense $^{14}$N(p,$\gamma$)$^{15}$O resonance at 278 keV. Because these background reactions emit gamma rays which are visible in the BGO prompt gamma-ray spectrum, the event counts in the corresponding peaks were used to constrain the yields of \nuc{13}{N} and \nuc{15}{O} in the fit. Figure \ref{fig:ActivationSpectrum} shows a representative annihilation-rate plot.

\begin{figure}[!htb]
    \centering
    \includegraphics[width=1.0\linewidth]{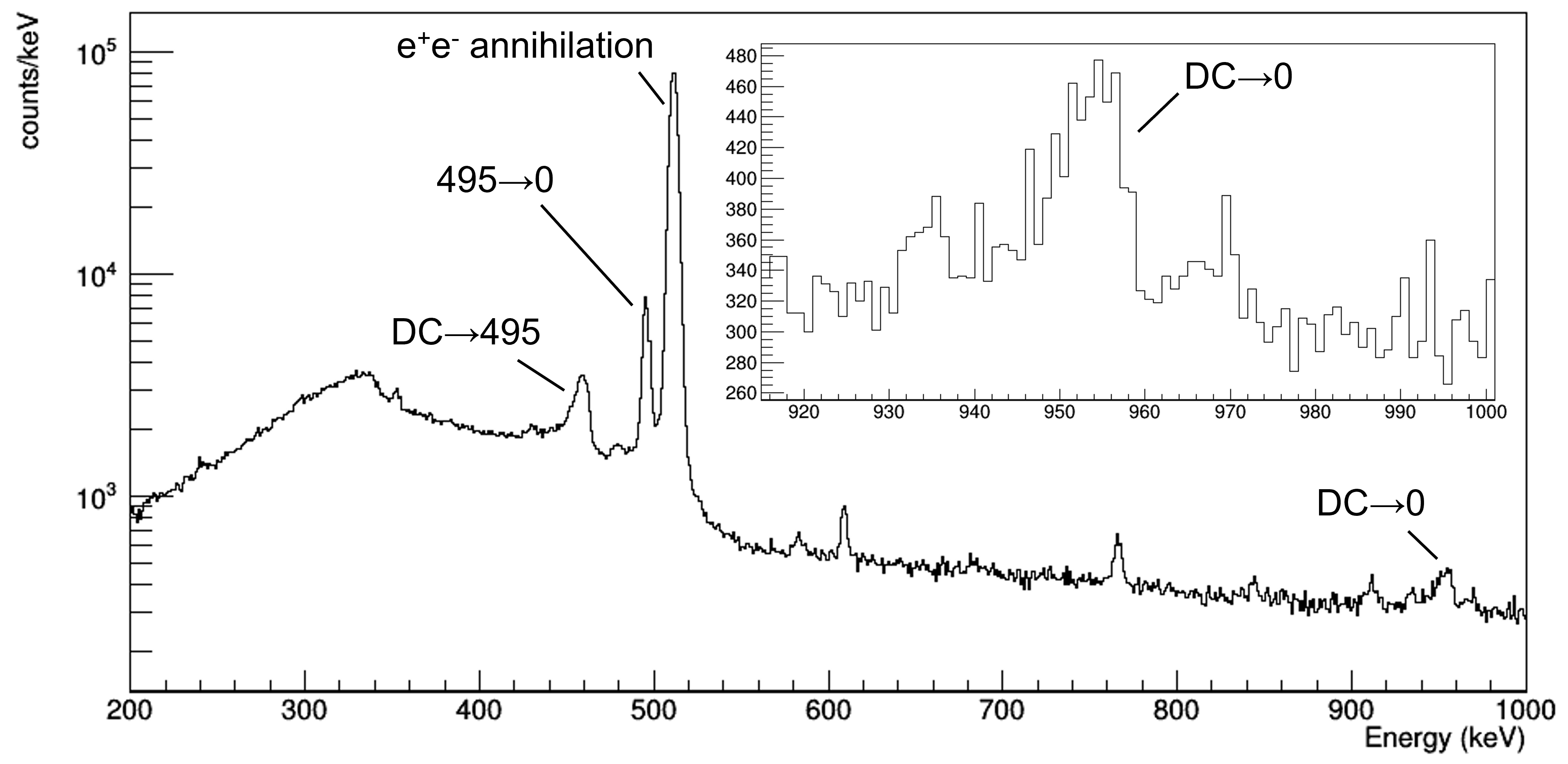}
    \caption{HPGe spectrum obtained at \Ep=380~keV. The \dctog{} peak is shown in the inset. The peaks at 583 keV, 609 keV, 768 keV and 911 keV stem from the decay chains of environmental $^{238}$U and $^{232}$Th.}
    \label{fig:PromptSpectrum}
\end{figure}

For the prompt gamma-ray experimental campaign, the reaction yield %, i.e. the number of reactions per incident beam particle \cite{Iliadis-book}, 
was calculated from the number of observed counts from each of the two primary transitions. Figure \ref{fig:PromptSpectrum} shows a sample spectrum from the HPGe detector. % obtained at a beam energy of 380~keV.
%In the \CeBr{} spectra the peaks were gaussian to a good approximation. However, in the HPGe spectra the peaks were fitted with the following empirical equation, as in \cite{ChilleryThesis}:

%\begin{equation}
%    \label{eq:DCFunction}
%    f(E_{\gamma})=\frac{Ae^{-2\pi\eta}}{E_{\rm{CoM}}}(1+\rm{e}^{\frac{E_{\rm{fit}}-E_{\gamma}}{\delta R}})^{-1}(1+\rm{e}^{\frac{E_{\gamma}-E_{\rm{fit}}-\Delta E}{\delta L}})^{-1}
%\end{equation}

%In most cases a linear background was appropriate, but in the HPGe spectra with \Ep{} $<$ 260 keV, the \dctof{} peak lay on the Compton shoulder of the 511 keV positron annihilation peak. It was found that the Compton shoulder could be described by an error function with a shape that did not strongly depend on proton energy. 

The yields from each detector were corrected for the angular distribution factor $W(\theta) = 1 + \sum\limits_{n=1}^{4}\{Q_{n}A_{n}P_{n}\left[cos(\theta)\right]\}$, where $P_{n}$ are the Legendre polynomials of order $n$ and the terms $Q_n$ are attenuation coefficients due to the finite solid angle covered by each detector, evaluated from Geant4 simulations. The angular distribution coefficients $A_n$ were derived by extrapolating the coefficients from Ref. \cite{Chow1975} to the relevant beam energies. The validity of the extrapolation was verified by comparing the yields from the three detectors at E$_{\textrm{cm}}$ 370~keV, where the gamma-ray peaks from the capture to the first excited state and to the ground state were clearly visible in all three detectors. %The $S$ factors for the \dctog{} and \dctof{} transitions were then summed to give the total $S$ factor.
More details on the analysis will be provided in a forthcoming publication \cite{Robb-inprep}.

The astrophysical $S$ factor was calculated as \cite{Brune2013} %using the effective energy method \cite{Brune2013} by the equation:

\begin{equation}
    \label{eq:EffectiveEnergySFactors}
    %S(E_{\textrm{cm}})=Y(E_{\textrm{cm}})\epsilon_{\rm{eff}}(E_{\textrm{cm}})\biggl(\int_{E_{\textrm{cm}}-\Delta E}^{E_{\textrm{cm}}}\frac{e^{-2\pi\eta} P(E)}{E} dE\biggr)^{-1},
    S(E_{\textrm{cm}})=Y\biggl(\int_{E_{0}-\Delta E}^{E_{0}}\frac{e^{-2\pi\eta} P(E)}{\epsilon_{\rm{eff}}(E) E} dE\biggr)^{-1},
\end{equation}

\noindent where $\epsilon_{\rm{eff}}$ is the effective stopping power of the target calculated using SRIM-2013 stopping powers \cite{SRIM} and $P(E)$ is the distribution of \nuc{16}{O} atoms as a function of the depth inside the target, determined with the NRRA technique and corrected at each beam energy for the different stopping power. The results are shown in Figure \ref{fig:TotalSFactors} and provided in tabular form in the Supplemental Material \cite{supplemental_1}, with statistical and systematic uncertainties reported separately. Statistical uncertainties on the activation data are below 2.5\%. On the other hand, the statistical uncertainty on the prompt-gamma data range from 1.0\% to 42\%, depending on the transition observed and the center-of-mass energy. Systematic uncertainties were evaluated independently for the prompt-gamma and activation experiments, and the systematic effects common to both techniques were identified, as summarized in Table \ref{tab:Systematics}. Specifically, the uncertainties on beam energy \cite{Formicola2003}, accumulated charge \cite{Gesue24-PRL} and effective stopping power (calculated from the mean errors quoted by SRIM \cite{SRIM}) are the same for the two experimental approaches. 

%The uncertainties on beam energy and target thickness effectively acted as modifications to the integration limits in equation \ref{eq:EffectiveEnergySFactors}. These effects were evaluated by varying the limits within their respective uncertainty ranges. 
The effects of the uncertainties on beam energy and target thickness were evaluated by varying the integration limits in equation \ref{eq:EffectiveEnergySFactors} within their respective uncertainty ranges. It was found that the beam energy uncertainty had negligible impact ($\ll1\%$), while the uncertainty due to target thickness depends on the precision achieved with the NRRA technique. The latter varied between the two experimental campaigns, due to the significantly different gamma-ray detection efficiencies of the setups used. %In the activation experiment the uncertainty on the target thickness was 1.0\%, while in the prompt gamma experiment it was 5\%, both resulting in comparable contributions to the uncertainty on the $S$-factor.
%\textcolor{red}{A statistical test was performed to check whether the apparent fluctuations in the data points could indicate additional structures, leading to the conclusion that the data are consistent with a non-resonant cross section.}
 
\begin{table}[h!]
\begin{center}
\begin{tabular}{|c|c|c|}
\hline
\textbf{Source} & \textbf{Activation} & \textbf{Prompt} \\
& & \textbf{Gamma}\\
\hline
Beam Energy & \multicolumn{2}{c|}{$\ll1\%$} \\
Beam Charge & \multicolumn{2}{c|}{$2\%$} \\
Effective Stopping Power & \multicolumn{2}{c|}{$3\%$} \\
Target Thickness & $\sim$1\% & $\sim$5\% \\
Detection Efficiency & 5\% & 3\% - 5\%\\
Angular Distributions & & 2.5\% \\
\hline
\end{tabular}
\end{center}
\caption{Systematic uncertainties on the $S$-factors}
\label{tab:Systematics}
\end{table}

%\begin{table}[h!]
%\begin{center}
%\caption{\textit{Systematic uncertainties adopted during the activation experiment $S$-factor calculations}}
%\label{tab:ActivationSystematics}
%\begin{tabular}{|c|c|c|}
%\hline
%\textbf{Source} & \textbf{Value} & \textbf{Commonality} \\
%\hline
%Beam Energy & 0.3\,keV & Common \\
%Accumulated Charge & 1.5\,\% & Common \\
%Effective Stopping Power & 3.0\,\% & Common \\
%Activation Efficiency & 5\,\% & Common \\
%Target Thickness & 1.0\,\% & Point-to-Point \\
%\hline
%\end{tabular}
%\end{center}
%\end{table}

%\begin{table}[h!]
%\begin{center}
%\caption{\textit{Systematic uncertainties adopted during the prompt gamma ray experiment $S$-factor calculations}}
%\label{tab:PromptSystematics}
%\begin{tabular}{|c|c|c|}
%\hline
%\textbf{Source} & \textbf{Value} & \textbf{Commonality} \\
%\hline
%Beam Energy & 0.3\,keV & Common \\
%Accumulated Charge & 1.5\,\% & Common \\
%Effective Stopping Power & 3.0\,\% & Common \\
%Propagated Angular Distributions & 2.5\,\% & Common \\
%Target Thickness & 5.1\,\% & Point-to-Point \\
%\hline
%\end{tabular}
%\end{center}
%\end{table}

%--------------------------------------------------------------
\section{\label{sec:fit} $S$-factor fit and reaction rate evaluation \protect\footnotemark}

%\section{\label{sec:results}Results}
The \react{16}{O}{\pg}{17}{F} reaction %has no resonances at low energies, therefore it 
occurs by direct capture in a single-step process \cite{ROLFS197329,CHRISTY196189,Iliadis2022}. Such direct processes are described accurately within a single-particle potential model, treating $^{16}$O and the proton as structureless, with their wave functions generated considering single-particle Woods-Saxon (WS) potentials. 
The radiative capture cross section is obtained by computing electromagnetic transitions between bound and scattering wave functions. In this work, we consider the E1 transition using the exact expression for the radial operator \cite{Rolfs1973}. We found, as in Ref.~\cite{Iliadis2022}, that E2 transitions are negligible and the reaction is driven by E1 transitions.  

In this two-body model, the $5/2^+$ ground state and $1/2^+$ excited state of $^{17}$F are seen, respectively, as a proton bound in a $d$- and an $s$-wave to the $0^+$ ground state of $^{16}$O. The depths of the WS potentials in these partial waves are adjusted to the experimental proton separation energies of 600.27(25)~keV for the $5/2^+$ ground state \cite{Wang2012} and 104.94(27)~keV for the $1/2^+$ excited state \cite{TILLEY19931}. The normalization of the bound-state wave functions
%, i.e. the spectroscopic factors 
and the radius of the WS potentials are adjusted to reproduce the $S$-factor data. As in Ref. \cite{Iliadis2022}, we have found that the radiative capture is mainly peripheral, i.e., the total $S$-factor is sensitive only to  the Asymptotic Normalization Coefficients (ANCs) of the ground states, especially at low energies in the range of astrophysical interest. %As in Ref. \cite{Iliadis2022},
%Because the radiative capture is a peripheral reaction, %i.e. its cross section scales only with the normalization of the tails of the bound-state wavefunctions, 
%we use the (ANCs) as free parameters.} %, as in Ref. \cite{Iliadis2022}. }
 These ANCs are determined in the single-particle potential model by the geometry of the  potentials generating the wave function, the binding energy of the states and the spectroscopic factors. 
Owing to the high correlation between the radius and diffuseness of the  potentials, we consider a fixed diffuseness $a=0.65$~fm in all partial-waves to avoid over-fitting.

Since the capture occurs mainly through E1 transitions, captures to the ground and excited states occur from scattering states in $p$ and $f$ waves. As in Ref.~\cite{Iliadis2022}, we found the capture at low energies ($<1$~MeV) to be insensitive to the choice of the potentials used to generate the scattering waves. In this work we use the scattering wave functions generated using a hard sphere potential in the $p$ and $f$ waves, because they reproduce accurately the energy dependence of the data at energies greater than $1$~MeV. 
The only adjustable parameters of the theoretical model are the ANCs of the bound state and the radii of the WS potentials. These parameters are constrained with the $S$ factor data measured in this work and datasets from Refs.~\cite{Morlock1997,Chow1975,Hester1958}. %, which reported separate statistical and systematic errors. 
The best-fit parameters and their uncertainties are obtained with a frequentist approach. This choice is motivated by the results of a recent Bayesian analysis of this reaction \cite{Iliadis2022}, which obtained posterior distributions exhibiting an approximate Gaussian shape. In such conditions, we expect that both Bayesian and frequentist approaches will lead to similar optimal parameters and uncertainties ~\cite{PhysRevC.110.064606,PhysRevC.111.035802}. %(see Refs.~\cite{PhysRevC.110.064606,PhysRevC.111.035802} for detailed comparisons). 

In a frequentist approach, %one determines the free parameters by minimizing a goodness-of-fit $\chi^2$ function. One has to carefully choose the form of this function, as there are many possible choices when considering multiple datasets~\cite{BARLOW2021164864,PhysRevC.111.035802,DAGOSTINI1994306} and it can influence the final 
the choice of the $\chi^2$ function to minimize can influence the final results~\cite{PhysRevC.110.064606,PhysRevC.111.035802}. We choose to scale the contribution of each data set to the $\chi^2$ by their number of data points, so that larger datasets do not dominate the fit. Separately, we introduced normalization factors for the datasets from Refs.~\cite{Morlock1997,Chow1975,Hester1958} to account for their systematic errors. The $\chi^2$ includes a penalty from the normalization factor, which is treated as a data point with a mean of 1 and an error equal to the systematic uncertainty of the associated dataset.
%We choose to normalize the contribution of each data set by their number of data points, to avoid larger datasets dominating the fit, and we introduce normalization factors for the datasets from Refs.~\cite{Morlock1997,Chow1975,Hester1958} to account for their systematic errors. \color{red} The normalisation factors introduce an additional penalty to the $\chi^2$ according to the systematic uncertainty of the associated dataset.\color{black}

%This $\chi^2$ function is defined as 
%\begin{equation}
%    \begin{split}
%        \chi^2 & = \frac{1}{N_0}\sum_{n}\frac{(S_{0,n}-S(\theta)_{0,n})^2}{\sigma_{stat,0,n}^2} +\\
%        & + \frac{1}{N_i}\sum_{i\neq 0} \left(\sum_n\frac{(f_iS_{i,n}-S(\theta)_{i,n})^2}{(f_i\sigma_{stat,i,n})^2}+\frac{(f_i-1)^2}{\sigma_{sys,i}^2}\right)\label{eq:costFct}
%    \end{split}
%\end{equation}

%where $S_{i,n}$ denotes the $n^{th}$ $S$ factor data  of the $i^{th}$ data set (with $i=0$ for the data obtained in this work), $\sigma_{stat,i,n}$ and $\sigma_{sys,i,n}$ correspond to the statistical and systematic uncertainties. 
%The uncertainties on the best-fit parameters are obtained from the covariance provided by the minimization procedure, rescaled by the best-fit $\chi^2_{min}/d$ with $\chi^2_{min}$ the value of the goodness-of-fit function for the best-fit parameters and $d$ the degrees of freedom.
    
% \color{red}[CH: I did not talk about the different energy dependence of the activation vs prompt data, that can be added if needed]\color{black}

\begin{figure}[!htb]
    \centering
    \includegraphics[width=0.8\linewidth]{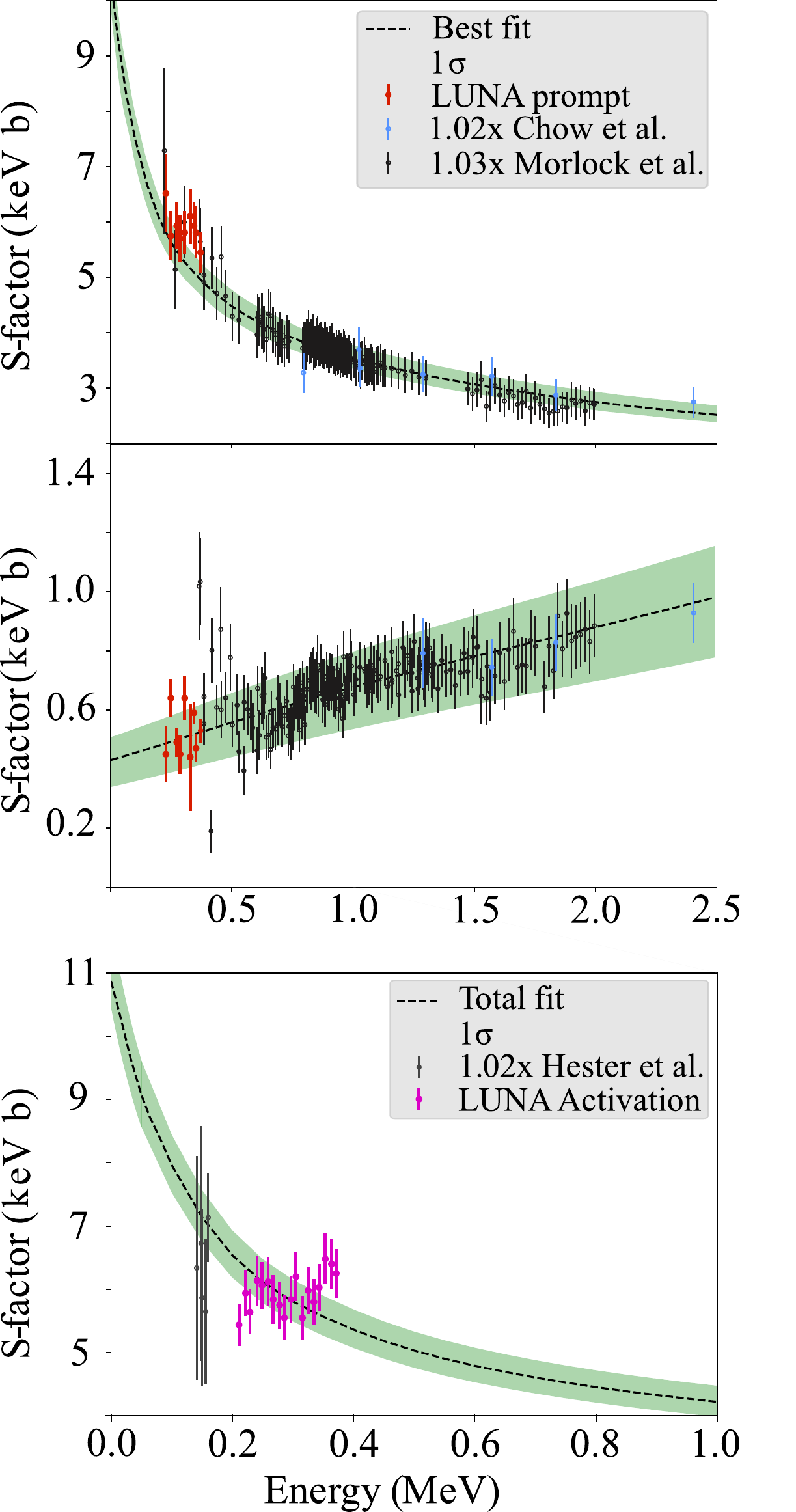}
    \caption{$S$ factors for the $^{16}$O$(p,\gamma)^{17}$F radiative capture reaction to the $1/2^+$ excited state of $^{17}$F (top), $5/2^+$ ground state of $^{17}$F  (middle) and the total $S$-factor (bottom). %The black data points correspond to the prompt gamma detection and activation measured in this work, the other data points are normalized according to the legend and are taken from~\cite{Morlock1997,Chow1975,Hester1958} (with data from Ref. \cite{Morlock1997} corrected  as in \cite{Iliadis2008}). 
    The error bars include both statistical and systematic errors. The black dashed lines are the evaluation obtained with the best-fit parameters and the green bands the corresponding $1\sigma$ uncertainties. To assess whether the observed fluctuations could indicate significant peak-like structures, we repeated the fits by adding a positive localized component with variable centroid and width to the smooth S-factor parameterization. The inclusion of this term did not lead to a statistically significant improvement of the fits in the energy region under discussion, and its fitted amplitude was compatible with zero within the uncertainties.}
    \label{fig:TotalSFactors}
\end{figure}

\begin{figure}[!htb]
    \centering
    \includegraphics[width=1.0\linewidth]{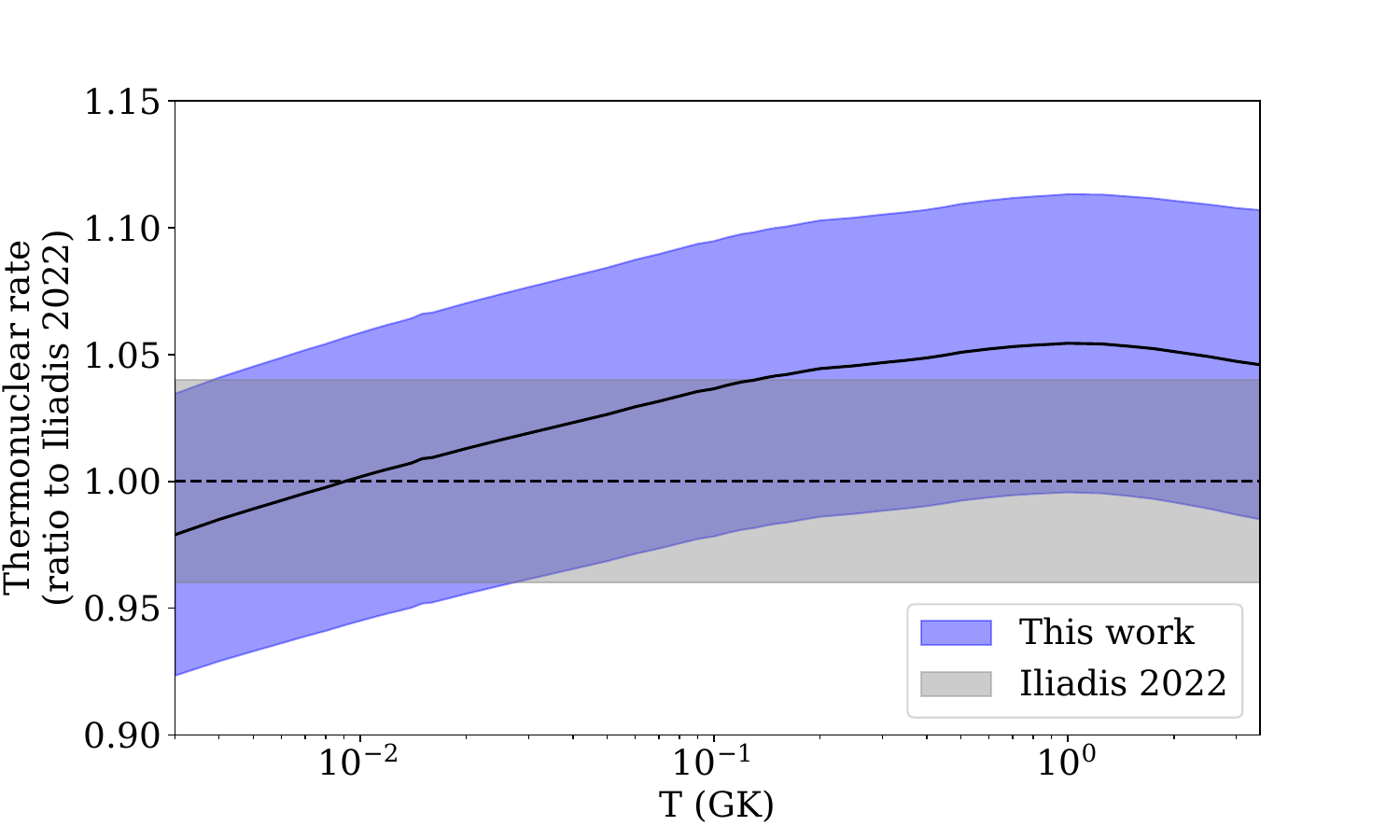}
    \caption{Ratio of the $^{16}$O$(p,\gamma)^{17}$F reaction rate obtained in this work (blue) with the evaluation of ~\cite{Iliadis2022} (grey). The bands represent the $1\sigma$ uncertainties.}
    \label{fig:Rate}
\end{figure}

Fig.~\ref{fig:TotalSFactors} shows the experimental $S$ factors for the transition to the the $1/2^+$ state, the $5/2^+$ state and the total $S$ factor. Because the fitted ANCs and potential depths for the two transitions are independent of each other, the $\chi^2$ takes into account simultaneously the quality of the fits of the two partial $S$-factors and the total $S$-factor, compared with the respective available datasets. Compared to other datasets, the prompt and activation data measured in this work lie at lower energies and exhibit reduced uncertainties. Including the activation data does not significantly change the overall fitted $S$-factor and its uncertainties. A future publication will discuss the difference between fitting the LUNA prompt data only or both prompt gamma and activation data. 
 %Using the statistical and theoretical mod}els described above, 
The resulting ANCs of the $1/2^+$ and $5/2^+$ states are $\mathcal{C}^{1/2^+}=80(2)$~fm$^{-1/2}$ and $\mathcal{C}^{5/2^+}=0.96(6)$~fm$^{-1/2}$, 
and the total $S$-factor at 0.01~MeV is 10.4(1.0)~keV~b \footnote{By construction, the uncertainties on the ANCs incorporate those of the spectroscopic factors.}. A comparison of the present results with those obtained by fitting only non-LUNA data and with literature fits is provided in the Supplemental material \cite{supplemental_2}. 
%The ANCs of the $1/2^+$ and $5/2^+$ states %are evaluated as $\mathcal{C}^{1/2^+}=80(4)$~fm$^{-1/2}$ and $\mathcal{C}^{5/2^+}=0.96(6)$~fm$^{-1/2}$, 
%and the $S$-factors at 0.01~MeV for transitions to those two states are given in the Supplemental material, together with a comparison of the present results with those obtained by fitting only non-LUNA data. %as $S^{1/2^+}(0.01)=10.1(1.0)$~MeV~b and $S^{5/2^+}(0.01)=0.35(5)$~MeV~b, leading to a total $S$-factor of $S(0.01)=10.4(1.0)$~MeV~b.
%The impact of the new LUNA data on the ANCs and $S$-factors is shown in Table \ref{tab:SummarySfactor}, where the results obtained by fitting only non-LUNA data are also reported.
%Interestingly, both prompt and activation data impact these evaluated quantities and their uncertainties, as repeating our procedure without including the new LUNA data leads to a smaller ANC and $S$ factor for the transition from the dominant $1/2^+$ state (compare the first two lines of Table~\ref{tab:SummarySfactor}). Moreover, 
The inclusion of our data in the evaluation leads to more precise $S$-factors, and ANCs that are closer to the values extracted from transfer reaction data on $^{16}$O \cite{PhysRevC.59.1149}.
The astrophysical reaction rate was evaluated using the fitted $S$-factor including all literature data. Figure~\ref{fig:Rate} shows a comparison of the present rate with the previous evaluation of Ref.~\cite{Iliadis2022}. Our evaluation exhibits larger uncertainties than \cite{Iliadis2022}, although it includes additional low-energy data, which constrain the low-energy behavior of the $S$ factor. This difference is %in part due to a small tension between these low-energy data and the previous datasets and 
mostly due to the use of different statistical and theoretical frameworks: Ref.~\cite{Iliadis2022} uses a Bayesian approach that incorporates prior knowledge, resulting in a different choice of likelihood function. %, e.g., which includes in a different manner the renormalization factors,
A detailed analysis considering various choices of likelihood functions and scattering potentials will be provided in a forthcoming publication~\cite{Robb-inprep}.

%\begin{figure}
%\includegraphics{fig_1}
%\caption{\label{fig:S-fac}Figure with $S$-factor results.}
%\end{figure}

\section{\label{sec:astro} Astrophysical implications}
The new \react{16}{O}{\pg}{17}{F} rate was used to explore if simultaneous measurements of stellar carbon and oxygen isotopic ratios can provide a new pathway to constrain the initial mass and the maximum temperature attained during mixing events. These parameters are interdependent, making their individual effects difficult to isolate.
%extent of mixing events in \textcolor{red}{first-ascent} RGB stars. \textcolor{red}{The stellar initial mass and the maximum temperature attained during mixing are interdependent, making their individual effects difficult to isolate.}
A good target for this kind of investigation is $\alpha$-Boo (Arcturus), the brightest star in the northern celestial hemisphere. %, with apparent visual magnitude $m_V=-0.5$.
Arcturus is a typical RGB star, with mass in the range of 1 to 2 M/M$_\odot$ (see \cite{abia2012}, and references therein). %The present day composition, as deduced from optical and IR spectra is summarized in table \ref{tab:arcturus}.
%The metallicity of Arcturus is clearly lower than solar ([Fe/H]$=-0.52$ \footnote{$[X/Y]=\log \frac{n_X}{n_Y}- \log \left ( \frac{n_X}{n_Y} \right )_\odot$, where $n_X$ is the density of the element X}), while oxygen is enhanced relative to iron ([O/Fe]$\sim 0.4$). Concerning the CNO isotopic composition, \cite{abia2012} report \ciso$=9\pm1$ and \oiso$=3030\pm530$. Such a low \ciso clearly indicates the activation of a deep-mixing process.
Reported carbon and oxygen isotopic ratios for this star are \ciso$=9\pm1$ and \oiso$=3030\pm530$ \cite{abia2012}. After the first dredge-up, the carbon isotopic ratio should be $\sim25(\pm5)$, where the quoted range account for variations of the initial mass and metallicity \cite{Boothroyd1999ApJ,straniero2017AA}; therefore, the substantially lower value measured for Arcturus proves the activation of some form of extra-mixing during the RGB phase.
% Such a low \ciso clearly indicates the activation of extra mixing, in addition to the first dredge-up.
By means of the FuNS code \cite{Straniero2019}, we computed a set of evolutionary tracks for stars of initial mass between 1 and 2 M$_\odot$ and chemical composition in agreement with available spectroscopic studies of Arcuturus, specifically, [Fe/H]$=-0.52$ and [O/Fe]$= 0.4$  \cite{smith2013}.  The adopted nuclear network includes 35 isotopes and 29 reactions, whose rates are from the latest release of the STARLIB database \cite{STARLIB}, except for the proton captures on \ele{12}{C} and \ele{13}{C}, for which we adopt the updated rates from \cite{c12c13_prl2023}, the \oopa and \oopg reactions from \cite{rapagnani_2025}, and the new \op rate presented in this paper. 

According to the extant spectroscopic analysis, a solar \ciso  (95.5 \cite{2025_lodders}) and an enhanced \oiso (7165 \cite{abia2012,smith2013}) were adopted as initial abundance ratios in all our models. After the first dredge-up, depending on the assumed initial mass, the \ciso and the \oiso ratios are reduced down to 23-27 and 500-6000,respectively. A further reduction can occur after the RGB bump, when the H-burning shell reaches the steep molecular-weight gradient left by the first dredge-up, and the extra-mixing switches on. We assume the extra-mixing to starts from the inner border of the convective envelope and extends down to a layer where the temperature attains T$_{\textrm{max}}$. A constant mixing velocity (v$_{\textrm{mix}}$) is assumed. In practice, this simplified description of the extra-mixing phenomenon depends on 3 input parameters: v$_{\textrm{mix}}$, T$_{\textrm{max}}$ and the initial mass (M$_i$). Actually, when v$_{\textrm{mix}}$ is varied between 0.01 and 0.0001 times the average turbulent velocity in the innermost layer of the convective envelope, both \ciso and \oiso change very little. In contrast, these isotopic ratios are sensitive to variations in T$_{\textrm{max}}$ and M$_i$.  

\begin{figure}[!htb]
    \centering
    \includegraphics[width=1.0\linewidth]{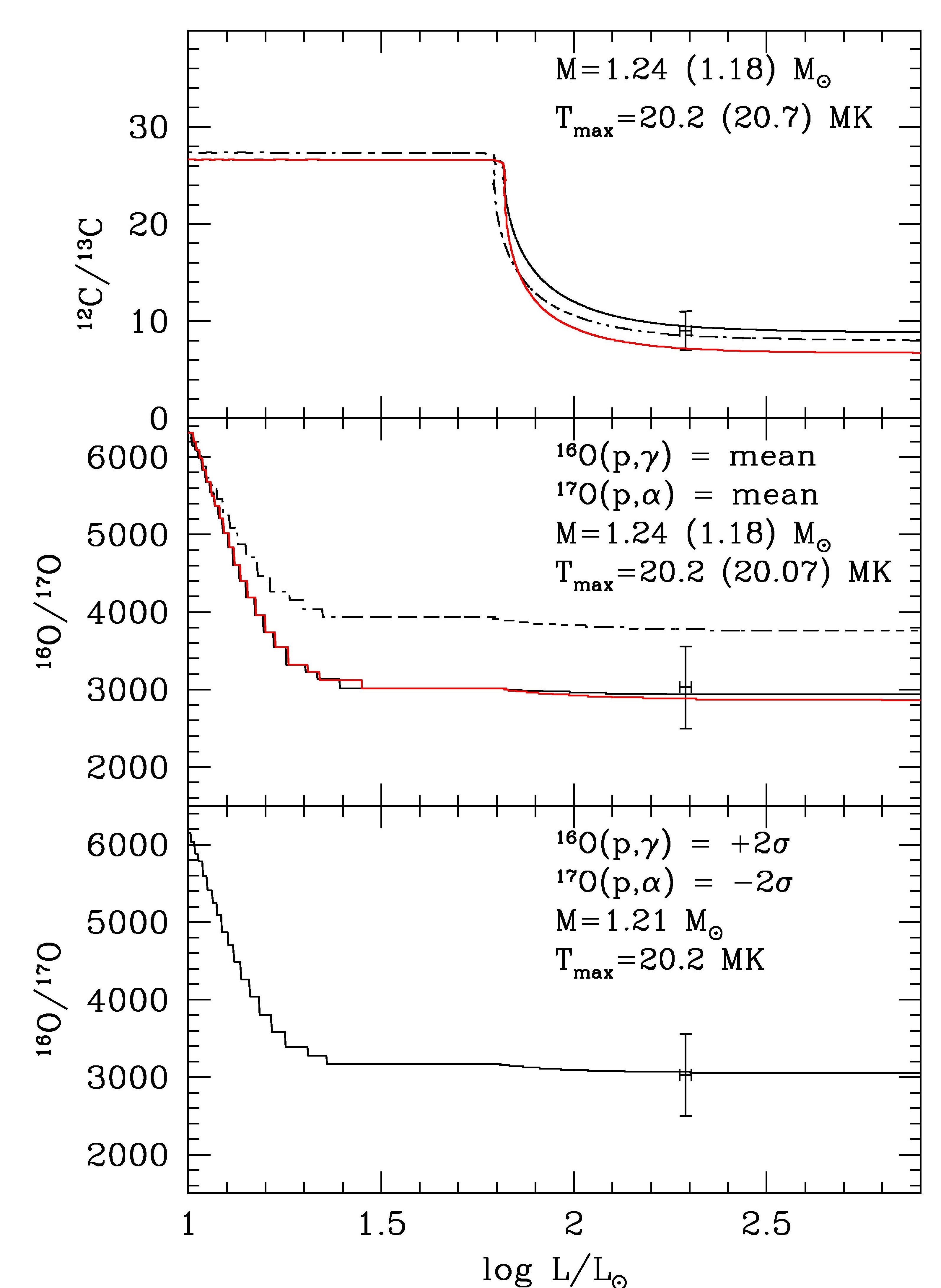}
    \caption{Best fits of the \ciso (top panel) and \oiso (middle and bottom panels) ratios of Arcturus as a function of the stellar luminosity. The ratios from spectroscopic observations are reported as points with error bars. In the top and middle panel, the black-solid and the black-dashed lines represent model predictions for $M_i=1.24$ and $1.18$, respectively, and T$_{\textrm{max}}=20.2$ MK, while the red line was obtained  assuming  $M_i=1.24$ and T$_{\textrm{max}}=20.7$ MK. The bottom panel shows a fit obtained increasing the \react{16}{O}{\pg}{17}{F} by 2$\sigma$ and decreasing the \react{17}{O}{\pa}{14}{N} rate by 2$\sigma$ with respect to the mean values. The corresponding best-fit mass is 1.21 M$_\odot$.}
    %Best fits of the \ciso (left) and \oiso (right) observed for Arcturus (squares with error bars) as a function of the stellar luminosity, obtained using different \react{16}{O}{\pg}{17}{F} and \react{17}{O}{\pa}{14}{N} reaction rates. The black-solid and the black-dashed lines in the low panels represent model predictions for $M_i=1.24$ and $1.18$, respectively, while the red line was obtained for $M_i=1.24$, but adopting the STARLIB rates for the \react{16}{O}{\pg}{17}{F} and \react{17}{O}{\pa}{14}{N} reactions. The O isotopic ratio is only marginally affected by the nuclear physics uncertainty, thus allowing a high-precision determination of the initial stellar mass of RGB stars and, in turn, of their ages.}
    \label{fig:arctaurus}
\end{figure}

As a result, the simultaneous fit of \ciso and \oiso, as shown in Figure \ref{fig:arctaurus}, provides stringent constraints for T$_{\textrm{max}}=20.2\pm0.5$ MK and M$_i=1.24\pm0.04$. 
Here, the quoted errors account for the uncertainties on the measured C and O isotopic ratios, while 
the rates of CNO reactions, specifically, the $^{12}$C$(p,\gamma)^{13}$N, $^{13}$C$(p,\gamma)^{14}$N, $^{14}$N$(p,\gamma)^{15}$O, \op and \oopa, imply a further 3\% uncertainty in the predicted M$_i$ and a 2\% in T$_{\textrm{max}}$. Noteworthily, a $\pm2\sigma$ variation of the \op reaction rate results in a change of less than 1\% in M$_i$, indicating that this rate is now well constrained.
%As a whole, the T$_{\textrm{max}}$ and M$_i$ uncertainties reported here are dominated by the \ciso and \oiso stellar measurements.

In summary, we presented new measurements of the \react{16}{O}{\pg}{17}{F} reaction at energies between 200 and 380~keV. Two complementary techniques were employed: prompt gamma-ray detection, yielding partial cross sections for capture to the ground and first excited states of \ele{17}{F}, and activation, which allowed us to determine the total cross section. A global fit of our data together with existing literature data was performed using a single-particle potential model. The resulting $S$-factor was used to derive an updated astrophysical reaction rate and to infer the initial mass and maximum temperature attained during deep-mixing events in the Arcturus RGB star, based on measured \ciso and \oiso ratios. We find that simultaneous measurements of these isotopic ratios can disentangle and place stringent constraints on M$_i$ and T$_{\textrm{max}}$, and that the uncertainty is now dominated by stellar abundance measurements. This technique could be applied in the future to any RGB star, provided that a precise spectroscopic abundance analysis is available.

%This result paves the way toward obtaining more precise constraints to the modeling of mixing processes in RGB stars.

\begin{acknowledgments}
We acknowledge funding from the following: INFN, The Hungarian National Research, Development, and Innovation Office (Grant No. NKKP ADVANCED 152233), ChETEC-INFRA (EU project no. 101008324), CNRS through the AIQI-IN2P3 project, the Helmholtz Association (VH-NG-22-08), LUNANOVA  (Eu project no. 101224374), the HUN-REN Network (ELKH-KMP 2023/4, KMP-2024/11), the Lendület Program LP2023-10 of the Hungarian Academy of Sciences.
E.M. acknowledges an Alexander von Humboldt post-doctoral fellowship.
\end{acknowledgments}

\bibliography{16O_bibliography}% Produces the bibliography via BibTeX.

\end{document}